\documentclass{aa}
\usepackage[dvips]{graphics}

\begin{document}

   \thesaurus{24     
              (01.2   ;  
               06.19.3)} 
   \title{Wavelet entropy as a measure of solar cycle complexity}

   \author{S. Sello
          }

   \institute{Mathematical and Physical Models, Enel Research,
              Via Andrea Pisano 120, 56122 Pisa - Italy\\
              email: sello@pte.enel.it
             }


   \maketitle

   \begin{abstract}

Using wavelet analysis approach, we can derive a measure of the \emph{disorder}
content of solar activity, following the temporal evolution
of the so-called \emph{wavelet entropy}. The interesting feature of this parameter
is its ability to extract a dynamical complexity information, in terms
of frequency distribution of the energy content, avoiding restrictions,
common in the nonlinear dynamics theory, such as stationarity.
The analysis is performed on the monthly time series of sunspot numbers.
From the time behaviour of the wavelet entropy we found a clear increase in the
disorder content of solar activity for the current $23^{th}$ solar cycle. This result
suggests general low accuracies for current solar cycle prediction methods.
Moreover, we pointed out a possible connection between wavelet entropy behaviour
and solar excursion phases of solar dipole.

   \end{abstract}

%

\section{Introduction}

Multiscale analysis, based on the wavelet approach, has been successfully used
for many different physical applications, including global solar activity
(Ochadlick et al. \cite{Ochadlick}; Lawrence et al. \cite{Lawrence}). Intermittence features have been well analysed
through the use of a continuous wavelet transforms on the monthly sunspot numbers
time series. More recently, different authors used longer daily sunspot numbers
time series to investigate more accurately new subtle periodicities and their
evolutions (Frick et al. \cite{Frick}; Ballester et al. \cite{Ballester}; Fligge et al. \cite{Fligge}; Sello \cite{Sello20}).
A recent work on wavelet applications to biomedical signals, shows the utility
of a new parameter, derived from wavelet formalism, to quantify the degree of
\emph{disorder} in the time series. This quantity, called \emph{wavelet entropy},
essentially gives an information on the scales extension involved in the energy
distribution, i.e. a dynamical complexity parameter (Quian Quiroga et al. \cite{Quian}).
The main aim of the present work is the application of the wavelet entropy
approach to the solar activity, through the wavelet analysis of the monthly
mean sunspot numbers.

\section{Wavelet entropy}

Fourier analysis is an adequate tool for detecting and quantifying constant periodic
fluctuations in time series. For intermittent and transient multiscale phenomena,
the wavelet transform is able to detect time evolutions of the frequency
distribution. The continuous wavelet transform represents an optimal localized
decomposition of time series, $x(t)$, as a function of both time $t$ and frequency
(scale) $a$, from a convolution integral:
\begin{equation}
  W(a,\tau)=\frac{1}{a^{1/2}} \int_{-\infty}^{+\infty} dt \quad x(t) \psi^ \star \left(\frac{t-\tau}{a} \right)
\end{equation}
where $\psi$ is called an analysing wavelet if it verifies the following
admissibility condition:
\begin{equation}
 c_{\psi}=\int_{0}^{+\infty} d \omega \quad \omega^{-1} \left \bracevert \hat{\psi} (\omega) \right \bracevert ^2 < \infty
\end{equation}
where:
\begin{equation}
 \hat{\psi} (\omega)= \int_{-\infty}^{+\infty} dt \quad \psi(t) e ^{-i \omega t}
\end{equation}
is the related Fourier transform. In the definition, $a$ and $\tau$ denote the dilation
(scale factor) and translation (time shift parameter), respectively.
We define the local wavelet spectrum:
\begin{equation}
 P_{ \omega} (k,t)=\frac{1}{2c_{ \psi} k_0} \left \bracevert W \left( \frac{k_0}{k},t \right) \right \bracevert ^2, \quad k \geq 0
\end{equation}
where $k_0$ denotes the peak frequency of the analysing wavelet $\psi$.
From the local wavelet spectrum we can derive a mean or global wavelet spectrum,
$P_{ \omega} (k)$:
\begin{equation}
 P_{ \omega} (k)=\int_{-\infty}^{+\infty} dt \quad P_{ \omega} (k,t)
\end{equation}
The relationship between the ordinary Fourier spectrum $P_{F} ( \omega)$ and the
 mean wavelet
spectrum $P_{ \omega} (k)$ is given by:
\begin{equation}
 P_{ \omega} (k)=\frac{1}{c_{ \psi} k} \int_{0}^{+\infty} d \omega \quad P_{F} (\omega) \left \bracevert \hat{\psi} (\frac{k_0 \omega}{k}) \right \bracevert ^2
\end{equation}
indicating that the mean wavelet spectrum is the average of the Fourier spectrum
weighted by the square of the Fourier transform of the analysing wavelet $\psi$
shifted at frequency k.
Here we used the family of complex analyzing wavelets consisting of a plane wave
modulated by a Gaussian, called Morlet wavelet, (Torrence et al. \cite{Compo}):
\begin{equation}
 \psi (\eta)=\pi^{-1/4}e^{i \omega_0 \eta} e^{- \eta ^2 /2}
\end{equation}
where $\omega_0$ is the non dimensional frequency here taken to be equal to 6 in
order to satisfy the admissibility condition, Eq.(2).
Following Quian Quiroga et al. (\cite{Quian}), we define a \emph{wavelet entropy} as a function of time:
\begin{equation}
 WS_t=- \sum_{k} p_{t,k} log_2(p_{t,k})
\end{equation}
where:
\begin{equation}
 p_{t,k}=\frac{P_{\omega}(k,t)}{\int dk P_{\omega}(k,t)}
\end{equation}
is the energy probability distribution for each scale level.
From the definition, follows that an ordered activity corresponds to a narrow
frequency distribution of energy, with low wavelet entropy, and a random activity
corresponds to a broad frequency distribution, with high wavelet entropy. Of
course, higher values for wavelet entropy means higher dynamical complexity,
higher irregular behaviour, lower predictability. The application of the wavelet
entropy is optimal for non-stationary signals.

\section{Solar activity: sunspot numbers}

We considered here the record of solar activity given by the monthly mean number
of sunspots, from SIDC archive (URL: http://www.oma.be/KSB-ORB/SIDC/index.html), covering the time interval: 1749-2000.04 and
consisting of 3013 observations.
Using a proper wavelet analysis we found the results displayed in Fig.1. Following
the notation described in Sello (\cite{Sello20}), the upper part shows the original time series in
its natural units. Time is here expressed in years. The central part shows the
amplitudes of the wavelet local power spectrum in terms of an arbitrary greyscale
contour map. White higher values are strong energetic contributions to power
spectrum, while black lower values are weak energetic contributions. Horizontal
time axis corresponds to the axis of time series and vertical scale (frequency)
axis is, for convenience, expressed in log values of cycles per year$^{-1}$. Thus
the range analyzed is between 148 years (value -5) and 134 days (value 1). The
right part shows the mean global wavelet power spectrum (solid line, SM) obtained with
a time integration of the local wavelet power spectrum, and the $5\%$ significance
level using a red noise autoregressive lag-1 background spectrum with
$\alpha=0.93$ (dotted line, SC) More precisely, we assumed here that different realizations of the considered stochastic process, given by sunspot activity, will be randomly distributed about a red noise background spectrum (i.e. increasing power with decreasing frequency), which is used to verify a \emph{null hypothesis} for the significance of a given peak in the wavelet power spectrum. In particular, if a peak in the wavelet power spectrum is significantly above the assumed red noise background spectrum, then it can be considered a real feature of the original signal with a given confidence level. This statement appears well justified from different applications to many complex physical signals, e.g. geophysical data. In order to obtain a simple model for red noise we considered a univariate lag-1 Markov process: $x_n=\alpha x_{n-1}+z_n$, where $\alpha$ is the assumed lag-1 autocorrelation, and $z_n$ is a Gaussian process. Note that: $\alpha=0$ corresponds to a white noise. (for more details see: Torrence et al. \cite{Compo}).
From the figure, it is possible to distinguish the main region corresponding to the well defined
Schwabe ($\sim 11$ years) cycle, with a dominant energetic contribution above the red
noise background spectrum. In particular, the complex
evolution of the cycle, both for scales and amplitudes, is well evident. The irregular finite
extension in the map represents a continuous interaction of frequencies which
generate complex dynamical behaviour. In order to better quantify the evolution
of complexity, or disorder for the solar cycle time series, we computed the time
evolution of the wavelet entropy. An interesting feature of this approach is that the
usual assumed hypothesis of stationarity of time series is here not necessary.

\begin{figure}
 \caption{Wavelet analysis of monthly mean sunspot numbers - right: mean global wavelet power spectrum (solid line, SM); and $5\%$ significance level (dotted line, SC)}.
 \label{lab1}
\end{figure}

Fig.2 (lower part) shows the time evolution of the wavelet entropy, Eq.(8).
For comparison, an ordered deterministic (harmonic) signal gives the numerical result displayed by the dashed line.
It is interesting to note that the degree of disorder reaches
maximum local values, often in phase with solar maxima, where the magnetic activity results
higher, but with many important exceptions (see, for example, cycles 20 and 22). The recorded maximum entropy value has been reached during the maximum of solar cycle 4 (1789). Moreover, the entropy results quite
low for long periods, including solar cycles from 5 to 7 (1798-1828) and, in
particular, for solar cycles from 9 to 14 (1848-1912). For the last three cycles, 21, 22 and
23, we found a clear average increase of the wavelet entropy, suggesting a more complex
dynamics, with a higher level of disorder due to a broad frequency energy
distribution, connected to evolution of the global magnetic activity. This
property forces a reduction in the accuracy level reachable by prediction methods, 
which are based on some strong deterministic hypothesis. For nonlinear chaotic dynamics methods
this means, at least, a lower value for Lyapunov's predictability time, (Sello \cite{Sello99}).
Furthermore, we note that the level of wavelet entropy can help to quantify the intermittence degree in solar activity caused by all the stochastic sources and, in particular, by stochastic fluctuations in meridional circulation of plasma, which are able to influence significant unpredictable variation in amplitude and phase of the solar cycle, as recently evidenced by Dikpati \& Charbonneau (\cite{Dikpati}).
Finally, we point out that the new information, added by the wavelet entropy, it is not useful for sunspot predictions.

\begin{figure}
 \caption{Monthly sunspot numbers (up) and wavelet entropy (down) time evolutions. The wavelet entropy for an ordered harmonic signal, is also displayed (dashed line).}
 \label{lab2}
\end{figure}

In order to investigate more accurately the periodicities which drive the evolution
of the wavelet entropy, we performed a wavelet analysis on the above derived time
series. For this analysis we used a wavelet software provided by C. Torrence and
G. Compo, and available at URL: http://paos.colorado.edu/research/wavelets.

\begin{figure}
 \caption{Wavelet analysis of the wavelet entropy - right: global power spectrum (solid line, POW); and $5\%$ significance level (dotted line, SL).}
 \label{lab3}
\end{figure}

Fig.3 depicts the wavelet map with $5\%$ confidence level regions. Here the scale axis
shows the periods in log values. The u-shaped line corresponds to the so-called 
Cone of Influence, i.e. the region of the wavelet spectrum in which edge effects, due to the finite-length of signal, become important. (Torrence et al. \cite{Compo}). The global power spectrum (right part, POW) indicates that
there is mainly one significant persistent periodicity related to the wavelet entropy, and localized around 40 years, a possible period doubling value of the main Schwabe cycle (Lawrence et al. \cite{Lawrence}). Other detected lower periodicities are more localized and energetically weaker, below the $5\%$ significance level of the assumed red-noise background spectrum (right part, SL).
The time evolution of the wavelet entropy suggests also the existence of
other longer modulations, not resolvable in the available time series.

It is interesting to point out possible relations existing between wavelet entropy behaviour and solar excursion phases, recently investigated by Mursula et al. (\cite{Mursula}). The excursion phases of the solar dipole are linked to the strong half solar rotation periodicity and arise when the heliospheric current sheet is flat and tilted. In particular, the authors found a clear connection between sunspot activity and the solar dipole tilt.
More precisely, the overall occurrence of excursions remained very weak during solar cycles from 9 to 14, where the solar dipole moment was lower with a high stability of the coronal evolution, and this is coherent with the lowest values of wavelet entropy.
Instead, during the most recent cycles (15-22) we detected largest excursions of the solar dipole, with a different cycle distribution, larger instabilities and a more disturbed heliosheet. This behaviour is again coherent with the overall increase of the wavelet entropy for the last cycles. The suggested correlation between wavelet entropy and solar excursion phases is here limited to a pure qualitative viewpoint. Further investigations, with a more quantitavive evaluation based on geomagnetic indices, are thus needed in order to support a reliable strict link between the degree of disorder in the sunspot activity, as measured by wavelet entropy, and the solar dipole mechanism.

\section{Conclusions}

From wavelet analysis approach, we considered a measure of the \emph{disorder} content
of solar activity, following the temporal evolution of the so-called \emph{wavelet entropy}.
The analysis is performed on the monthly time series of sunspot numbers
and shows an interesting  multiscale evolution of the wavelet entropy. In 
particular, we found a clear increase in the disorder content of solar activity
for the current $23^{th}$ solar cycle.
This result suggests general low accuracies reachable for current solar cycle 
prediction methods, which are based on some strong deterministic hypothesis.
Finally, we suggested a possible strict correlation between the wavelet entropy qualitative behaviour and the solar excursion phases of solar dipole, linked to coronal complexity evolution.

\end{document}